\documentclass[twocolumn]{aastex631}
\usepackage{gensymb}
\usepackage{hyperref}

\newcommand{\C}[1]{\textcolor{red}{\textbf{\MakeUppercase{#1}}}}
\newcommand{\uniform}[2]{$\mathcal{U}\,(#1,#2)$}
\newcommand{\normal}[2]{$\mathcal{N}\,(#1,#2)$}
\newcommand{\jeffrey}{$\mathcal{J}$}
\newcommand{\vsini}{$v$\,sin\,$i_\star$}
\newcommand{\vsinsin}{$\sqrt{v\,\text{sin}\,i_\star}$\,sin\,$\lambda$}
\newcommand{\vsincos}{$\sqrt{v\,\text{cos}\,i_\star}$\,cos\,$\lambda$}
\newcommand{\cospsi}{$\text{cos}\,\psi$}
\newcommand{\coslam}{$\text{cos}\,\lambda$}

\begin{document}

\title{An Obliquity Measurement of the Hot Neptune TOI-1694b}

\author[0000-0002-9305-5101]{Luke B. Handley}
\affiliation{Department of Astronomy, California Institute of Technology, Pasadena, CA 91125, USA}


\author[0000-0001-8638-0320]{Andrew W. Howard}
\affiliation{Department of Astronomy, California Institute of Technology, Pasadena, CA 91125, USA}

\author[0000-0003-3856-3143]{Ryan A. Rubenzahl}
\affiliation{Center for Computational Astrophysics, Flatiron Institute, 162 Fifth Avenue, New York, NY 10010, USA}

\author[0000-0002-8958-0683]{Fei Dai}
\affiliation{Institute for Astronomy, University of Hawai`i, 2680 Woodlawn Drive, Honolulu, HI 96822, USA}
\affiliation{Division of Geological and Planetary Sciences,
1200 E California Blvd, Pasadena, CA, 91125, USA}
\affiliation{Department of Astronomy, California Institute of Technology, Pasadena, CA 91125, USA}

\author[0000-0003-0298-4667]{Dakotah Tyler}
\affiliation{Department of Physics and Astronomy, University of California, Los Angeles, CA 90095, USA}

\author[0000-0001-7058-4134]{Rena A. Lee}
\altaffiliation{NSF Graduate Research Fellow}
\affiliation{Institute for Astronomy, University of Hawai`i, 2680 Woodlawn Drive, Honolulu, HI 96822, USA}

\author[0000-0002-8965-3969]{Steven Giacalone}
\altaffiliation{NSF Astronomy and Astrophysics Postdoctoral Fellow}
\affiliation{Department of Astronomy, California Institute of Technology, Pasadena, CA 91125, USA}


\author[0000-0002-0531-1073]{Howard Isaacson}
\affiliation{{Department of Astronomy,  University of California Berkeley, Berkeley CA 94720, USA}}



\author[0000-0002-5812-3236]{Aaron Householder}
\affiliation{Department of Earth, Atmospheric and Planetary Sciences, Massachusetts Institute of Technology, Cambridge, MA 02139, USA}
\affil{Kavli Institute for Astrophysics and Space Research, Massachusetts Institute of Technology, Cambridge, MA 02139, USA}


\author[0000-0003-1312-9391]{Samuel Halverson}
\affil{Jet Propulsion Laboratory, California Institute of Technology, 4800 Oak Grove Drive, Pasadena, CA 91109, USA}


\author[0000-0001-8127-5775]{Arpita Roy}
\affiliation{Astrophysics \& Space Institute, Schmidt Sciences, New York, NY 10011, USA}


\author[0000-0002-6092-8295]{Josh Walawender}
\affiliation{W. M. Keck Observatory, 65-1120 Mamalahoa Hwy., Kamuela, HI 96743, USA}

\begin{abstract}

We present spectral observations of the multiplanet host TOI-1694 during the transit of TOI-1694b, a 26.1 $M_\oplus$ hot Neptune with a 3.77-day orbit. By analyzing radial velocities obtained from the Keck Planet Finder, we modeled the Rossiter-McLaughlin effect and constrained the sky-projected obliquity to ${9\degree}^{+22\degree}_{-18\degree}$, which is strong evidence for a nearly aligned orbit. TOI-1694b is one of fewer than ten small planets accompanied by confirmed outer giant planets for which the obliquity has been measured. We consider the significance of the outer planet TOI-1694c, a Jupiter-mass planet with a 1-year orbit, and its potential role in influencing the orbit of TOI-1694b to its current state. Incorporating our measurement, we discuss the bifurcation in hot Neptune obliquities and present evidence for an independent polar population. The observed polar planets nearly ubiquitously have periods of $\le 6$ days and mass ratios of $10^{-4}$. Early perturbations by outer companions from resonance crossings in the disk-dispersal stage provide the most compelling explanation for this population. Systems which lack the necessary configuration will retain their primordial obliquity, since hot Neptunes lack the angular momentum needed to realign their hosts on relevant timescales.
\end{abstract}


\section{Introduction} \label{sec:intro}

The stellar obliquity angle is among the most dynamically rich observable quantities for exoplanet systems. It describes the misalignment of a stellar host's spin axis with a planet's orbital angular momentum. Standard planetary formation models like core accretion (\cite{pollack1996}, \cite{bodenheimer2000}, \cite{hubickyj2005}) and protoplanetary disk gravitational collapse (\cite{boss1997}, \cite{michikoshi2007}) predict planets to form in orbital planes nearly parallel to the surrounding disk plane. Given a quiescent formational evolution, the expected stellar obliquity angle, denoted by $\psi$, should be near zero (an ``aligned'' system) due to a common net angular momentum between the accreting star and disk. The measurement of nonzero obliquity (misalignment) is a signpost of a dynamically active formation or evolutionary environment. A common way to measure the sky-projection of this angle, denoted by $\lambda$, is through the time-dependent radial velocity (RV) signature during a planetary transit called the Rossiter-McLaughlin (RM) effect (\cite{rossiter1924}; \cite{mclaughlin1924}). 

Larger planets near their host stars are the most amenable to RM measurements, and the observed obliquity distribution of these planets is by far the most well studied. Most notably is the correlation between hot Jupiter obliquities and stellar temperature -- hot stars with $T_\mathrm{eff} \gtrsim 6$,200~K are often found to house misaligned orbits, while cooler stars are more often found to be in aligned orbits (the ``$\lambda - T_\mathrm{eff}$" relationship, see \cite{winn2010}, \cite{albrecht2022}). Stars below this temperature possess convective envelopes which may enhance tidal orbit realignment processes due to their turbulent eddies \citep{zahn2008}. The lack of such envelopes in hotter stars \citep{pinsonneault2001} is thought to help explain the prevalence of misaligned hot Jupiters orbiting the hottest stars \citep{albrecht2012}.

These realignment effects, however, may also work to obscure our view of the primordial distribution of stellar obliquity angles around cool stars. While $\psi$ was initially interpreted as a pure measure of dynamical excitement, our observation of the $\lambda - T_\mathrm{eff}$ trend suggests that stellar properties (i.e., tides) play a critical role. We have some theoretical grapple on these tidal processes in relation to binary stellar systems \citep{mazeh2008}, but their application to exoplanetary systems remains naive without a detailed understanding of the internal structure of each planet \citep{ogilvie2014}. Assuming an equilibrium tide model, where spin–orbit alignment and orbital decay are parametrized by the same tidal quality factor, an estimate of the realignment timescale for convective stars was given by \cite{zahn1977} as

\begin{equation} \label{eq:synctime}
    t_\mathrm{sync} \sim \left( \frac{M_p}{M_\star} \right)^{-2} \left( \frac{a}{R_\star} \right)^{6} \textrm{years},
\end{equation}
where $M_p/M_\star$ is the planet-to-star mass ratio, and $a/R_\star$ is the ratio of orbital semimajor axis to stellar radius. For a 4 Jupiter mass planet on a 2 day orbit, this timescale might be $\sim$1~Gyr. Therefore, when the orbit of a massive planet is confirmed to be aligned, it may be unclear whether this indicates a quiescent evolution, or the primordial obliquity angle being erased in real time by tides \citep{winn2015}. The dependence of this timescale on mass ratio is critical, and motivates a push to smaller planets. For a body of roughly Neptune's mass, this timescale will be hundreds of times longer (far more than the age of the system). Therefore, measurements of the stellar obliquity for Neptune mass planets are more directly indicative of the underlying dynamics. 

It remains to be seen if our understanding of the observed eccentricities and obliquities of hot Jupiters are adequate in explaining that of hot Neptunes. While much rarer than their short period Jupiter counterparts (the `Neptune Desert'; \cite{szabo2011}, \cite{mazeh2016}), early models of hot Neptune mass loss by stellar irradiation proposed that these two planet types formed by the same mechanism long before their obliquities could be measured \citep{baraffe2005}. Preferential occurrence of these populations both around metal-rich stars and in single-transiting systems further support a shared origin \citep{dong2018}, and recent work shows that small planets also tend to have high obliquities around hot stars (\cite{louden2024}). As the census of Neptune obliquities increases, we can seek to improve our general understanding of orbital architecture as a function of time, stellar effective temperature, and planetary mass. 

This paper focuses on TOI-1694 (TIC-396740648), a K-type star ($T_\mathrm{eff}=5135,V=11.4$, \cite{mistry2023}) which hosts one such hot Neptune planet TOI-1694b ($R_p = 5.44\pm0.18 R_\Earth$, $M_p = 26.1\pm2.2 M_\Earth$) on a 3.77 day orbit. This system stands out because it also houses an eccentric ($e = 0.180\pm0.048$) distant giant companion, TOI-1694c ($M_c\,$sin$\,i = 1.05\pm0.05 M_J$, $P_c = 389.2\pm3.9$d) which was confirmed by RVs in \cite{VanZandt2023}. Outer giant companions are thought to contribute to the apparent abundance of misaligned systems through several processes. For example, von Zeipel-Kozai-Lidov oscillations (see \cite{wu2003}, \cite{fabrycky2007}, \cite{naoz2011}, and \cite{naoz2016}) can produce an array of orbital inclinations, including retrograde orbits. Non-planar orbits may also be excited in distant giant systems via resonance sweeping with the dispersing protoplanetary disk \citep{petrovich2020} or the evolving stellar hosts oblateness \citep{batygin2016}.

In any case, these systems are a dynamical laboratory for probing planetary formation, migration, and evolution. Our analysis of TOI-1694b provides an additional datapoint. In Section \ref{sec:transit}, we modeled transits of TOI-1694b to significantly improve the transit ephemeris. In Section \ref{sec:RM}, we modeled the RM effect induced during a transit of TOI-1694b using radial velocities from the Keck Planet Finder. In Section \ref{sec:dynamics}, we consider an aligned sky-projected obliquity under the context of several dynamical theories. Sections \ref{sec:discussion} and \ref{sec:conclusion} are a discussion of the broader context of stellar obliquity measurements in relation to small planets, and a conclusion.



\section{Transit Analysis} \label{sec:transit}

The precise determination of obliquity by modeling time-series RVs (Section \ref{sec:RM}) requires precise knowledge of the transit ephemeris, which we modeled using photometry from multiple sources.

\subsection{TESS Observations}

TOI-1694 was observed at 2 minute cadence during two concurrent Sectors of the Transiting Exoplanet TESS \citep{ricker2015} photometry -- Sectors 19 and 20 -- spanning 23 November 2019 UT to 20 January 2020 UT. A summary of the stellar and planetary parameters determined by previous authors is given in Table \ref{tab:stellarparams}. We queried the flux time series using the \textsc{lightkurve} package \citep{lightkurve} in \textsc{Python}, and adopted the Presearch Data Conditioning Simple Aperture Photometry (PDCSAP) values from the Science Processing Operations Center (SPOC) pipeline \citep{jenkins2016} for each observation. Neglecting gaps in the observing cadence during these Sectors, TESS observed 12 transits of TOI-1694b with a baseline of 49 days between the first and last event. This baseline is insufficient in constraining the transit ephemeris of TOI-1694b for obliquity measurements several years later. 


\begin{table}[]
    \centering
    \begin{tabular}{c|c|c}
         \hline
         Parameter & Value & Reference \\
         \hline
         $T_\textrm{eff}$ (K) & 5135$\pm$50 & A \\
         log $g$ (dex) & 4.658$\pm$0.100 & A \\
         $[M/H]$ (dex) & 0.06$\pm$0.08 & A \\
         $R_\star$ $(R_\odot)$ & 0.8183$\pm$0.0477 & A \\
         $M_\star$ $(M_\odot)$ & 0.84$\pm$0.03 & B \\
         \vsini\ (km~s$^{-1}$) & 1.2$\pm$1.0 & B \\
         \hline
         $M_b$ $(M_\Earth)$ & 26.1$\pm$2.2 & B \\
         $P_b$ (d) & 3.770179$\pm$0.000060 & A \\
         $R_b$ $(R_\Earth)$ & 5.5$\pm$0.5 & A \\
         $e_b$ & 0 & B \\
         \hline
         $M_c\,\text{sin}\,i_c$ $(M_\Earth)$ & 334$\pm$16 & B \\
         $P_c$ (d) & 389.2$\pm$3.9 & B \\
         $e_c$ & 0.18$\pm0.048$ & B \\
         \hline
    \end{tabular}
    \caption{Parameters for the TOI-1694 system by reference. \textit{Key:} A: \cite{mistry2023}, B: \cite{VanZandt2023}.}
    \label{tab:stellarparams}
\end{table}

TOI-1694 was again imaged by TESS in Sector 73, spanning 7 December 2023 UT through 3 January 2024 UT, but was only partially captured on the outermost pixel of the imaging area. The pipeline did not produce two-minute cadence products due to the diminished quality of the photometry. Instead, we queried the Sector 73 Full Frame Image (FFI) time series products which were produced at 200 seconds cadence using \textsc{tesscut} \citep{tesscut}. We identified the center pixel of TOI-1694, and used the python package \textsc{unpopular} \citep{unpopular} to detrend systematics (i.e., pointing jitter and scattered light from the Earth) from the flux time-series. \textsc{unpopular} uses a Causal Pixel Model (CPM) approach \citep{wang2017} to model these systematics in a data-driven manner, which was crucial given the difficulty of conducting aperture photometry at the detector edge. After subtracting the CPM signal from the light curve of the center pixel in Sector 73, the transits of TOI-1694b became visible.


Next, we masked observations within a transit duration (2.851 hours) on either side of the projected transit midpoints as calculated from the transit posteriors in \cite{mistry2023}. We fit cubic splines to the out of transit flux to identify any systematic relics, and divided the Sector 73 time series residuals by the splines to flatten it. The light curve was then suitable for comparison with the previous Sectors.

For reasons we develop in Appendix~\ref{sec:dragonfly}, we also analyzed a ground-based followup observation using the Dragonfly multi-lens array \citep{abraham2014}, but we chose to model only the TESS photometry.

\subsection{Transit Model} \label{sec:transitmodel}

We used the 4-year baseline between observations with TESS in Sectors 19, 20, and 73 to tightly constrain the transit parameters of TOI-1694b. Sectors 19 and 20 were treated as a single continuous dataset. We assumed the measurement uncertainties of the Sector 73 flux to be constant in time but unknown, and elected to estimate their underlying distribution using a penalty term alongside a standard chi-squared likelihood function

\begin{equation}
    -2 \,\text{ln}\, \mathcal{L} = \chi^2_{19,20} + \sum_{i}\left( \left( \frac{y_{73}(t_i) - f(t_i)}{\sigma_{\text{73}}} \right)^2 + \text{ln}\, 2\pi\sigma_{\text{73}}\right),
\end{equation}
where $\mathcal{L}$ is the model likelihood, $\chi^2_{19,20}$ is the usual chi-squared statistic for the data in Sectors 19 and 20, $y_{73}$ is our derived FFI flux values, and $\sigma_{73}$ is the unknown flux error in those measurements. We generated the true light curve model $f(t)$ using the package \textsc{batman} \citep{batman}. The full list of parameters adopted is shown in Table \ref{tab:rmparams}. We assumed a quadratic limb darkening model parametrized by the coefficients $q_1$ and $q_2$ as described in \cite{Kipping2013}. We set the prior distributions to be Gaussian with means determined by \textsc{exofast} \citep{exofast}, and with width of 0.3. For the dynamical quantities, we use uninformative uniform priors which surround the highest posterior density regions in \cite{mistry2023} by roughly $10\sigma$. We assumed zero eccentricity for TOI-1694b as in \cite{VanZandt2023}, and use a Jeffrey's prior for the jitter term $\sigma_{73}$.

\begin{figure}
    \centering
    \includegraphics[width=\linewidth]{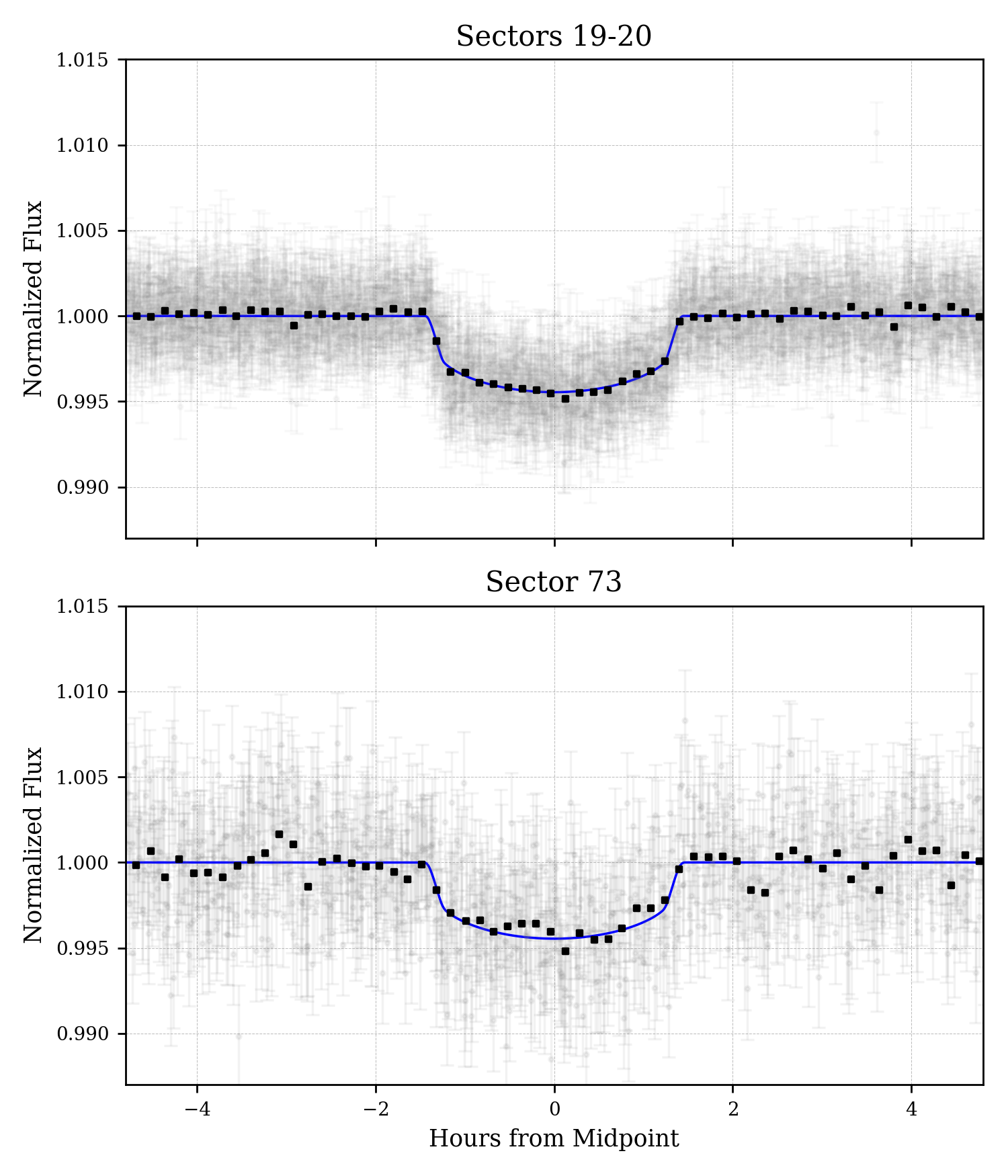}
    \caption{TESS light curves for TOI-1694b, phase folded to the median period of the MCMC posterior. PDCSAP light curves from the 120-second cadence observations in Sectors 19 and 20 are shown on top. Our derived light curves from 200-second cadence FFI observations in Sector 73 are below. The model is shown in blue, and the black squares are binned data points at 10 minute intervals.}
    \label{fig:transitfits}
\end{figure}

We computed the maximum a posteriori (MAP) fit to the data using the \textsc{python} package \textsc{scipy} \citep{scipy} with Powell's method \citep{powell1964} as the minimization routine. This solution was used as the starting vector of a Markov Chain Monte Carlo (MCMC) exploration of the posterior using the \textsc{emcee} package \citep{emcee}. We initialized 32 walkers and ran for 30,000 steps each to constrain the transit parameters. We checked that the Gelman-Rubin statistics of all parameters were below 1.01 as our convergence diagnostic. The results of the MCMC are shown in Table \ref{tab:rmparams} as medians of the posterior distributions along with the bounds of the 68\% confidence interval. The phase folded fit to each dataset (PDCSAP light curves versus our FFI light curves) is shown in Figure \ref{fig:transitfits}.

\section{Rossiter Mclaughlin Measurement} \label{sec:RM}

\subsection{KPF Radial Velocities} \label{sec:rvdata}

We observed TOI-1694 during a transit of TOI-1694b on 23 January 2024 using the Keck Planet Finder (KPF) spectrograph \citep{gibson2024} on Keck 1 in the normal read mode (47s readout time). The 
R$\,\sim\,$98,000 spectra were reduced using the standard KPF Data Reduction Pipeline\footnote{\url{https://github.com/Keck-DataReductionPipelines/KPF-Pipeline}} (DRP), as developed by members of the California Planet Search (CPS, \cite{howard2010}). The DRP computes radial velocities (RVs) by cross-correlating a template mask with the spectra across two color channels. We obtained 600s exposures, and the reduced 1D spectra achieved a peak of roughly S/N$\,\sim\,$110 in the green channel and S/N$\,\sim\,$130 in the red channel. We obtained 25 spectral observations which spanned a total of 4.5 hours, roughly 83 minutes of observations pre-ingress and 16 minutes of observations post-egress given the transit duration of 171 minutes. Table \ref{tab:rvdata} reports the instantaneous RV measurements (with a fiducial Keplerian model subtracted) and the associated uncertainties. We suspected the uncertainties were overestimated due to the developmental stage of the DRP. As we show in Section \ref{sec:model}, the true RV uncertainty is $\sim$1.41~m~s$^{-1}$, lower than all values in Table \ref{tab:rvdata}. In the following analysis, we chose to set all uncertainties to zero, and allow a jitter term $\sigma_\mathrm{RV}$ to inflate them equally.

\begin{table}[ht]
    \centering
    \begin{tabular}{|c|c|c|c|}
        \hline
        BJD - 2457000 & RV (m~s$^{-1}$) & $\sigma_\mathrm{RV}$ (m~s$^{-1}$) & \textbf{$t_\mathrm{exp}$} \\
        \hline
        3332.75261 & -0.40 & 1.64 & 600s \\
        3332.76026 & -0.01 & 1.56 & 600s \\
        3332.76791 & -1.75 & 1.56 & 600s \\
        3332.77542 & -0.20 & 1.46 & 600s \\
        3332.78257 & 0.57 & 1.51 & 600s \\
        3332.79020 & -0.67 & 1.55 & 600s \\
        3332.79779 & 1.51 & 1.52 & 600s \\
        3332.80512 & 3.82 & 1.56 & 600s \\
        3332.81305 & 2.59 & 1.59 & 600s \\
        3332.82024 & 2.82 & 1.60 & 600s \\
        3332.82766 & 2.76 & 1.57 & 600s \\
        3332.83538 & 4.14 & 1.61 & 600s \\
        3332.84293 & 2.47 & 1.55 & 600s \\
        3332.85042 & 3.90 & 1.5 & 600s \\
        3332.85772 & 1.01 & 1.46 & 600s \\
        3332.86517 & 0.37 & 1.58 & 600s \\
        3332.87840 & -3.54 & 1.72 & 600s \\
        3332.88602 & -4.20 & 1.84 & 600s \\
        3332.89346 & -3.49 & 2.03 & 600s \\
        3332.90117 & -1.06 & 1.96 & 600s \\
        3332.90944 & -3.82 & 1.91 & 600s \\
        3332.91709 & 0.12 & 1.79 & 600s \\
        3332.92533 & -0.18 & 1.71 & 600s \\
        3332.93264 & -0.24 & 1.73 & 600s \\
        3332.94025 & -1.40 & 1.71 & 600s \\
        \hline
    \end{tabular}
    \caption{RV measurements of TOI-1694 during transit, produced by the KPF DRP, after subtracting a linear Keplerian model. The time values are given as the Barycentric Julian Date (BJD) in the Barycentric Dynamical Time standard. The given RV errors are likely overestimated given our analysis in Section~\ref{sec:RM}.}
    \label{tab:rvdata}
\end{table}

\subsection{RM Model} \label{sec:model}

We modeled the RM effect induced by the transiting planet using the prescription of \cite{hirano2011}. We adjust the macroturbulent velocity in our model using the scaling relation in \cite{valentifischer2005} with the effective temperature in Table \ref{tab:stellarparams}. We adopt nominal values such as those in \cite{rubenzahl2021} for the natural line width at 1~km~s$^{-1}$ and microturbulent velocity at 0.7~km~s$^{-1}$. The instrumental broadening profile was calculated using the KPF line spread function. We neglected differential rotation due to the low rotational velocity of the star (a \vsini\ of 1.5~km~s$^{-1}$ would induce differential rotational modulation of the signal at only the 10~cm~s$^{-1}$ level), but included the effects of convective blueshift.

Our analysis centers on the retrieval of the sky-projected stellar obliquity angle $\lambda$, and the projected rotational velocity \vsini, which we sampled in the basis $\{$\vsincos,\vsinsin$\}$. We included an RV offset parameter $\gamma$, an RV trend parameter $\dot{\gamma}$, the convective blueshift velocity $v_\mathrm{cb}$, and the RV noise term $\sigma_{RV}$ as discussed in Section \ref{sec:rvdata}. We also include the transit parameters discussed in Section \ref{sec:transit} with the exception of the period and the time of conjunction, which were precisely determined. In total, our RM model was a function of 11 free parameters: $\{\lambda$, \vsini, $\gamma$, $\dot{\gamma}$, $v_\mathrm{cb}$, $\sigma_\mathrm{RV}$, $b$, $a/R_{\star}$, $R_p/R_\star$, $q_1$, $q_2\}$. 

We used another penalized chi-squared log likelihood to evaluate the performance of the RM model at each observation time.

\begin{equation}
    -2 \,\text{ln}\, \mathcal{L} = \sum_{i}\left( \left( \frac{\text{RV}(t_i) - g(t_i)}{\sigma_\mathrm{RV}} \right)^2 + \text{ln}\, 2\pi\sigma_\mathrm{RV}\right),
\end{equation}
where RV$(t_i)$ is the KPF RV at time $t_i$, and $g(t_i)$ is the model RV from \cite{hirano2011}. We oversampled the data by a factor of 5 and binned down in order to replicate the integrated effect over the 600-second exposures. We enforced a uniform prior on $\lambda$ across all angles, and another uniform prior on \vsini\ between 0~km~s$^{-1}$ and 5~km~s$^{-1}$. A Jeffrey's prior was used for the jitter $\sigma_{RV}$. For the convective blueshift velocity, we set a wide Gaussian prior of width 200~m~s$^{-1}$ centered on the expected value of -126~m~s$^{-1}$ from the solar scale factor prescription in \cite{liebing2021} using $T_\mathrm{eff}$. For the five transit parameters, we used our posteriors in Section \ref{sec:transitmodel} as priors by employing a gaussian kernel density estimation routine in the \textsc{scipy.stats} library. 

\begin{figure}
    \centering
    \includegraphics[width=\linewidth]{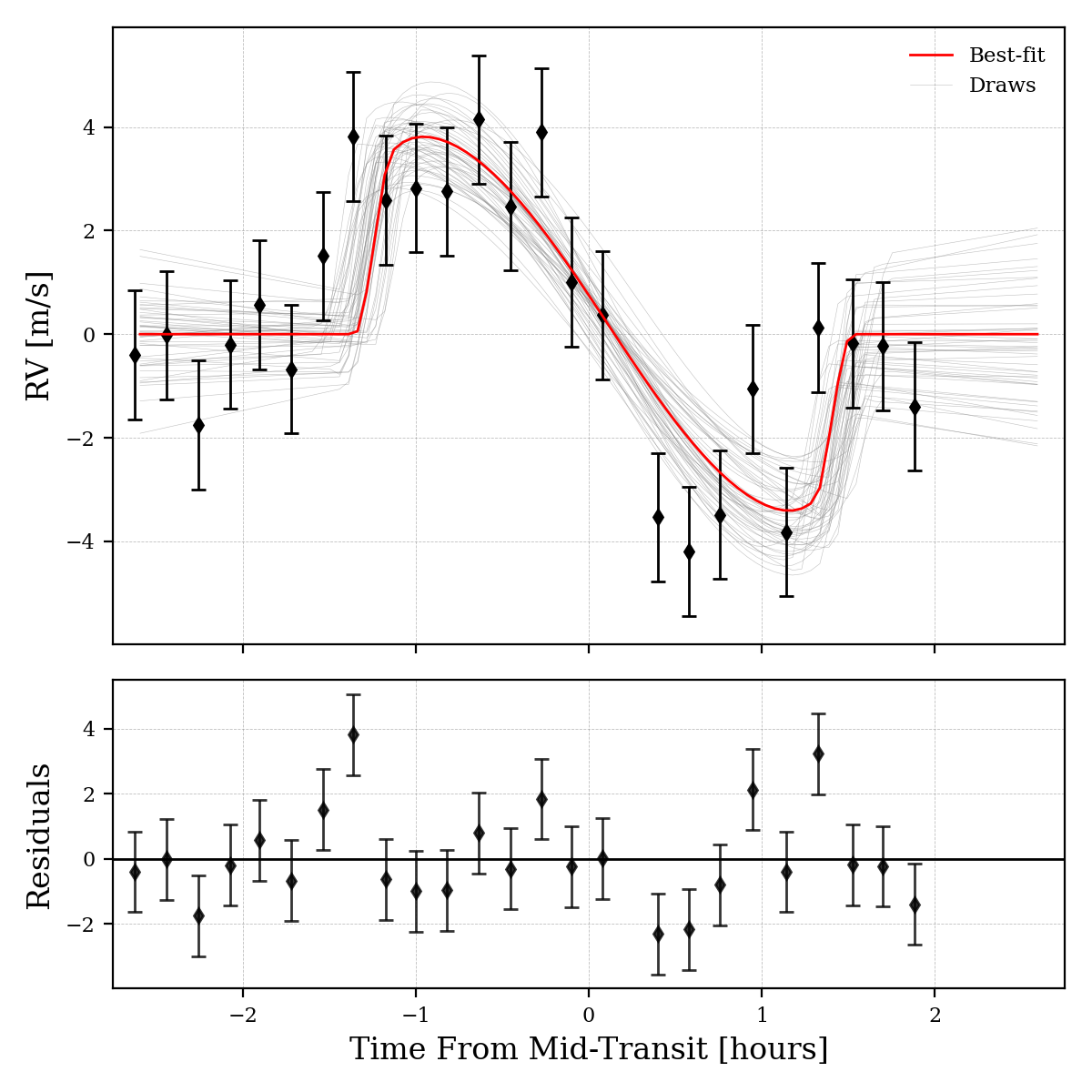}
    \caption{Our fit to the RM effect induced by TOI-1694b (red) compared to the observations in Table \ref{tab:rvdata} (black diamonds). Gray lines are random posterior draws from the MCMC chains.}
    \label{fig:RM}
\end{figure}

After computing the MAP fit, we initialized an MCMC for the RM model with 32 walkers for 12,000 steps. We discarded the first 1,000 steps as burn-in, and thinned the chains by a factor of five to get the best mixture of data. We confirmed that the Gelman-Rubin statistic was below 1.01 for all 11 parameters. A summary of the posterior distributions is given in Table \ref{tab:rmparams}, and a plot of the best fit solution can be found in Figure~\ref{fig:RM}. We detect evidence of an aligned orbit at $\lambda={9\degree}^{+22\degree}_{-18\degree}$, and a \vsini\ of $1.50^{+0.30}_{-0.26}$~km~s$^{-1}$, which is in good agreement with the spectral broadening estimation of $1.2\pm1.0$~km~s$^{-1}$ from \cite{VanZandt2023} but much more tightly constrained. We found convective blueshift to be negligible, returning only our prior distribution. Removing the effect altogether had no impact on the posterior distributions of $\lambda$ or \vsini.

\begin{table*}[]
    \centering
    \begin{tabular}{|c|c|c|c|}
        \multicolumn{4}{c}{\textbf{Model Parameters}} \\
        \hline
        Parameter & Prior & Reference & Posterior \\
        \hline
        $P_b$ (d) & \uniform{3.769579}{3.770779} & A & $3.7701379^{+3.2e-06}_{-3.3e-06}$ \\
        $T_c$ (BJD - 2457000) & \uniform{1817.2602}{1817.2722} & A & $1817.2664^{+0.0004}_{-0.00041}$ \\
        $\lambda$ (deg) & \uniform{-180}{180} & - & $9.27^{+22.15}_{-18.15}$ \\
        \vsini\ (km~s$^{-1}$) & \uniform{0}{5} & B &  $1.50^{+0.30}_{-0.26}$ \\
        $\gamma$ (m~s$^{-1}$) & - & - &  $0.01^{+0.63}_{-0.64}$ \\
        $\dot{\gamma}$ (m~s$^{-1}$ d$^{-1}$) & - & - &  $-0.26^{+6.11}_{-6.11}$ \\
        $v_{cb}$ (m~s$^{-1}$) & \normal{-126}{200} & C & $-71.75^{+195.09}_{-196.92}$ \\
        $\sigma_{RV}$ (m~s$^{-1}$) & \jeffrey & - & $1.41^{+0.25}_{-0.20}$ \\
        $\sigma_{73}$ (ppt) & \jeffrey & - & $2.96^{+0.02}_{-0.02}$ \\
        $b$ & \uniform{0}{1} & - &  $0.349^{+0.173}_{-0.188}$ \\
        $a/R_\star$ & \uniform{2.206}{15.206} & A & $10.11^{+0.52}_{-0.86}$ \\
        $R_p/R_\star$ & \uniform{0.046}{0.076} & A & $0.061^{+0.002}_{-0.001}$ \\
        $q_1$ & \normal{0.37182}{0.3} & D  & $0.383^{+0.145}_{-0.106}$\\
        $q_2$ & \normal{0.32033}{0.3} & D  & $0.572^{+0.199}_{-0.185}$ \\
        \hline
        \multicolumn{4}{c}{\textbf{Derived}} \\
        \hline
        $i_{orb}$ (deg) & - & - & $87.61^{+1.07}_{-0.99}$ \\
        $u_1$ & - & - &  $0.72^{+0.162}_{-0.181}$ \\
        $u_2$ & - & - & $-0.083^{+0.258}_{-0.217}$ \\
        \hline
    \end{tabular}
    \caption{Priors and posteriors for parameters in the transit model (TESS) and RM model (KPF). \uniform{x}{y} denotes a uniform distribution between $x$ and $y$. \normal{x}{y} is a normal distribution with mean $x$ and standard deviation $y$. \jeffrey$\,$ denotes a Jeffrey's prior. The parameter $i_{orb}$ is the orbital inclination of planet TOI-1694b. Coefficients $u_1$ and $u_2$ are the standard limb darkening coefficients given by the transformation in \cite{Kipping2013}. \textit{Reference Key:} A: \cite{mistry2023}, B: \cite{VanZandt2023}, C: \cite{liebing2021}, D: \cite{exofast}.}
    \label{tab:rmparams}
\end{table*}

We were unable to constrain the the true obliquity without a measurement of the stellar inclination. We attempted to model the rotation period with an Autocorrelation Function (ACF) method as implemented in \textsc{spinspotter} \citep{holcomb2022}, as well as a generalized Lomb-Scargle periodogram method \citep{zechmeister2009} using the methodology described in \cite{bowler2023} on the TESS photometry. However, the TESS light curves do not appear modulated by stellar rotation above the level of systematic noise, so we could not constrain the stellar inclination. Furthermore, assuming an edge-on inclination and our \vsini, we compute a rotation period of $38^{+8}_{-7}$ days, potentially not resolved within the TESS sector baselines.

\section{Obliquity Dynamics} \label{sec:dynamics}

In this section, we utilize our measurement of an aligned $\lambda$ to constrain the dynamical history of TOI-1694b. If the timescale of obliquity realignment of the star by TOI-1694b is long compared to the age of the system, e.g., that of Equation~\ref{eq:synctime}, then the planet's current obliquity is indicative of a primordial or ongoing dynamical event. Evaluating Equation \ref{eq:synctime} for TOI-1694b gives an approximate realignment time of $t_\mathrm{sync}\sim 10^{14}$ years, hence, the aligned obliquity of TOI-1694b is a product of dynamical history rather than tidal dissipation. 

Other hot Neptune planets with distant giant companions, such as WASP-107b (\cite{dai2017}, \cite{rubenzahl2021}) and HAT-P-11b (\cite{winn2010b}, \cite{hirano2011a}), are known to have highly oblique orbits. Several theories suggest that these obliquities arise from the presence of a distant giant planetary companion. Could these same dynamics be at play in the TOI-1694 system? We now analyze the system in the context of several of those processes, also enumerated in \cite{rubenzahl2021}, which we will revisit in Section \ref{sec:demographicdynamics} alongside the greater population of hot Neptunes.

\subsection{Transition Disk Resonance with TOI-1694c}

\cite{petrovich2020} showed that polar orbits can are generated for small planets with an outer giant companion by a resonance encountered during the disk-dispersal stage. In this model, the inner Neptune experiences nodal precession due to the stellar quadrupolar field, while the outer Jovian precesses due to the remaining disk mass external to its $\gtrsim$ 1 au orbit. As the disk depletes, the precession rate of the Jovian may drop to commensurability with that of the inner planet. Depending on relative strength of the stellar quadrupole to the planet-planet interactions (their $\eta_\star$) and the strength of General Relativity (GR) to the planet-planet interactions (their $\eta_{\text{GR}}$), a resonance crossing may result in polar orbits nearly 100\% of the time.

We now apply this model to TOI-1694. The dominant uncertainty in this model is that of the Pre Main Sequence (PMS) stars stellar quadrupole moment $J_2$, which depends strongly on the PMS stellar radius $R'_\star$ and rotation period $P'_\star$ as \citep{sterne1939}

\begin{equation}
    J_2 \propto \frac{k_2}{M_\star}\frac{{R'_\star}^3}{{P'    _{\star}}^2},
\end{equation}
which can evolve by several orders of magnitude. Following \cite{petrovich2020}, we assumed the tidal Love number $k_2 \approx 0.2$ \citep{claret2012}, and we adopt the stellar mass to be $0.84\pm0.03 M_\odot$ from Table \ref{tab:stellarparams}. We adopt generous uncertainties on $P'_\star$ by drawing from a uniform distribution between 2 days and 15 days, which spans the regime for K-type stars with disks (\cite{rebull2020}, \cite{nofi2021}). We estimate the stellar radius post planet formation ($\sim$10 Myr) given the stellar mass using both the PARSEC \citep{bressan2012} and BCAH15 \citep{baraffe2015} isochrones with $[M/H] = 0.06$ to assess the systematic uncertainty between models. This resulted in $R'_\star = 1.049\pm0.022 R_\odot$ from the PARSEC model and $R'_\star = 1.087\pm0.017 R_\odot$ for the BCAH15 model. We chose to adopt a conservative $R'_\star = 1.07\pm0.03 R_\odot$ to make both models consistent at the $1\sigma$ level, implying $J_2$ ranged roughly between $10^{-5}$ and $10^{-4}$.

We assumed the same transitional disk model as in \cite{petrovich2020} with a decay timescale of $\tau_\mathrm{disk}=$ 1 Myr and evaluate the final inclinations of TOI-1694b relative to the disk plane based on the criteria in their analysis. TOI-1694b can achieve several obliquities based on the relative importance of $\eta_\star$ and $\eta_{\text{GR}}$. In resonance, the longitude of the ascending nodes of both planets will evolve together as $\Omega_b - \Omega_c = \pi$. As a result, the inner planet can be shown to evolve towards $\text{cos}\,i_b = 0$, i.e. a polar orbit, barring excitation of the inner planet's eccentricity at high inclinations due to the outer companion. 

Qualitatively, if the stellar quadrupole is strong, it will cause $\Omega_b$ to precess at a rate much higher than that of the outer disk on $\Omega_c$. Since the precession rate of $\Omega_c$ can only decrease in time due to the decaying mass of the disk, resonance between the two angles becomes impossible. Further, if the resonance is indeed encountered, GR must play a strong role to achieve a polar orbit, as it shields the planet from eccentricity excitation (e.g. a ZKL resonance) and allows the inclination to become fully excited. We summarize possible end-states for TOI-1694b with the following four Cases:

\begin{enumerate}
    \item The stellar quadrupole dominates the two-planet interactions ($\eta_\star\gg 1$). The resonance is missed altogether, and TOI-1694b remains in an \textit{aligned orbit}.
    \item The resonance is crossed, but the adiabatic time (Equation 7 in \cite{petrovich2020}) is long compared to the speed with which the disk disperses. The precession rate of $\Omega_c$ drops too quickly for the resonance to be maintained, causing a \textit{minor inclination} of $10\degree-40\degree$ (see \cite{quillen2006}).
    \item The resonance is crossed in the adiabatic limit, but GR does not completely dominate ($\eta_{\text{GR}} < \eta_\star + 6$). The planet is broken out of the inclination growth early with an \textit{oblique orbit} ($60\degree-80\degree$) due to eccentric instability and retains residual eccentricity.
    \item The resonance is crossed adiabatically and GR dominates eccentric instability, guaranteeing a circular and \textit{polar orbit}.
\end{enumerate}

\begin{figure*}
    \centering
    \includegraphics[width=\linewidth]{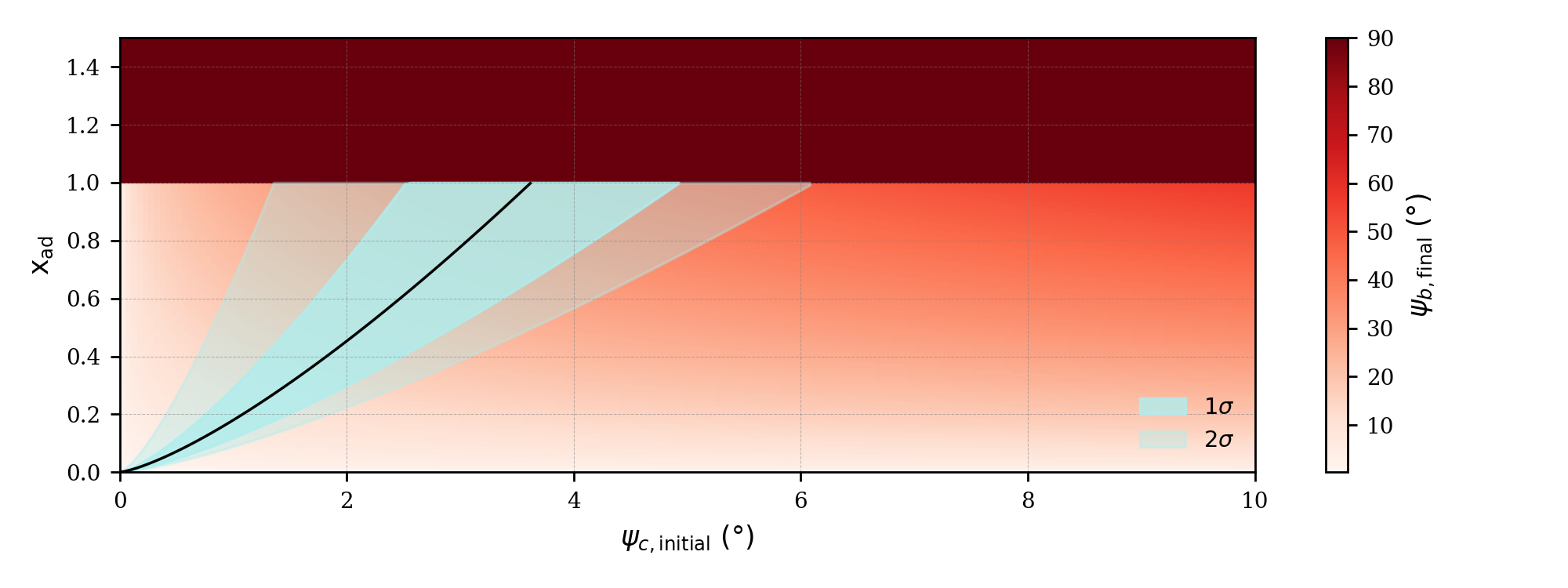}
    \caption{Most probable final obliquity of TOI-1694b for different inclinations of the outer companion assuming a resonance crossing. Regions of the parameter space which are allowed by both the model and our constraints on $\lambda$ are shaded in blue, with associated 1$\sigma$ and 2$\sigma$ confidence intervals. Near-aligned orbits are probable at low mutual inclinations.}
    \label{fig:diskmodel}
\end{figure*}

Incorporating all the relevant uncertainties on the PMS stellar parameters and the outer companion results in an $82.6\%$ probability of crossing the disk dispersal resonance. The adiabaticity of the crossing is strongly a function of the inclination of the giant companion TOI-1694c to the protoplanetary disk. We summarize the most probable obliquities as a function of the outer companions inclination $\psi_c$ in Figure \ref{fig:diskmodel}. If the outer companion is primordially inclined by $\psi_c \gtrsim 6\degree$, an adiabatic crossing is \textit{guaranteed} if the resonance is encountered. Since our measurement of $\lambda$ rules out an adiabatic crossing as high confidence (cases 3 and 4, above), a highly inclined companion is unlikely. However, an initial inclination of $\psi_c \lesssim 6\degree$ permits crossings of the resonance outside the adiabatic regime. After ruling out Cases 3 and 4 and integrating over small values of $\psi_c$, this results in a $51.6\%$ chance of a Case 1 orbit and a $48.4\%$ chance of a Case 2 orbit. 

In other words, the disk resonance model predicts a minor obliquity when the outer planet is nearly aligned. The posterior peak of $\lambda\sim 10\degree$ as well as any obliquity $\lesssim 40\degree$ are consistent with this model. A highly oblique orbit is only predicted for TOI-1694b for a misaligned outer planet. Thus, given our data, this model predicts a low mutual inclination between the two planets, i.e. TOI-1694c should also have a near-aligned stellar obliquity.

\subsection{Obliquity Oscillations}

The periodic evolution of mutual inclination between inner and outer planets may also dictate the distribution of hot Neptune obliquities. While our posterior distribution of $\lambda$ rules out a polar orientation, a $\sim50\degree$ orientation is permitted at the $2\sigma$ level. Assuming that the giant companion is nearly aligned with the stellar equator, the mutual inclination of the planets could surpass the minimum inclination ($i=39.2\degree$, \cite{kozai1962}, \cite{innanen1997}) to enter the Zeipel-Kozai-Lidov (ZKL) regime. During ZKL resonance cycles, a small body in a hierarchical three body system can exchange angular momentum with the secondary body (here, TOI-1694c), thereby oscillating synchronously between a highly eccentric orbit and a highly inclined one \citep{perryman2012}.  However, \cite{fabrycky2007} and  \cite{liu2015} showed that ZKL oscillations are quenched when additional short range effects (those not captured by the field of the distant companion, e.g. GR, tides, or rotational bulges) induce significant pericenter precession of the inner planets orbit, thereby disrupting the resonance. For the case of TOI-1694b, it is trivial to show that the GR apsidal precession timescale is roughly equal to that of ZKL \citep{fabrycky2007}:

\begin{equation} \label{eq:kozaigr}
    \frac{\tau_{\text{GR}}}{\tau_{\text{ZKL}}} = \left( \frac{\pi P_b^8}{4 G^2 M_\star^5} \right)^{1/3} \left( \frac{M_c c^2}{2 P_c^2} \right) \frac{\left( 1-e_b^2 \right)}{\left( 1-e_c^2 \right)^{3/2}} \approx 0.97
\end{equation}
Using $M_c \approx M_c\, \text{sin}\, i_c$. For values of $\tau_{\text{GR}}/\tau_{\text{ZKL}} \le 2.1$, \cite{fabrycky2007} showed that ZKL cycles fail to excite the inclination of the inner planet. A value of $\approx$ 2.1 is possible only if the Jovian's mass is several times higher. This would imply an initial mutual inclination between the two planets of $|\psi_b-\psi_c| \gtrsim 60\degree$ based on our viewing geometry. The null eccentricity of TOI-1694b further disfavors active ZKL cycles, although past cycles cannot be ruled out. If TOI-1694b formed on a longer period orbit, ZKL cycles followed by orbital decay through tidal friction could have conceivably deposited the planet into its present-day GR dominated configuration.

\cite{yee2018} showed that in the case of ZKL suppression, the obliquity may still evolve due to nodal precession. The longitude of ascending node $\Omega_b$ precesses about the invariant plane defined by the outer planets orbit, which dominates the total angular momentum of the system. We refer the reader to \cite{yee2018}, \cite{xuan2020}, and \cite{rubenzahl2021} for a more complete description in the context of obliquity evolution. Nodal precession causes oscillations in the obliquity of TOI-1694b between $\psi_b \in [0,2\psi_c]$. TOI-1694b may appear to have an aligned or highly misaligned orbit depending on the phase, but only if TOI-1694c is also substantially misaligned. Furthermore, TOI-1694b would preferentially be observed in a highly misaligned orbit \citep{xuan2020} if this were the case. 

Both ZKL and nodal precession can only drive large oscillations in the obliquity of TOI-1694b if the mutual inclinations with the outer planet is high. While a large obliquity of TOI-1694c can in principle be generated by stochastic phenomena such as planet-planet scattering \cite{}, this scenario further requires fortunate timing in obliquity cycles to be observed in its current state. Furthermore, an early scattering giving way to a large $\psi_c$ would result in a high adiabatic constant and therefore polar configuration (Figure \ref{fig:diskmodel}) in the disk-resonance framework.

\section{Discussion of Small Planet Obliquity Demographics} \label{sec:discussion}

\subsection{The Hot Neptune Obliquity Dichotomy}

\begin{figure*}
    \centering
    \includegraphics[width=\linewidth]{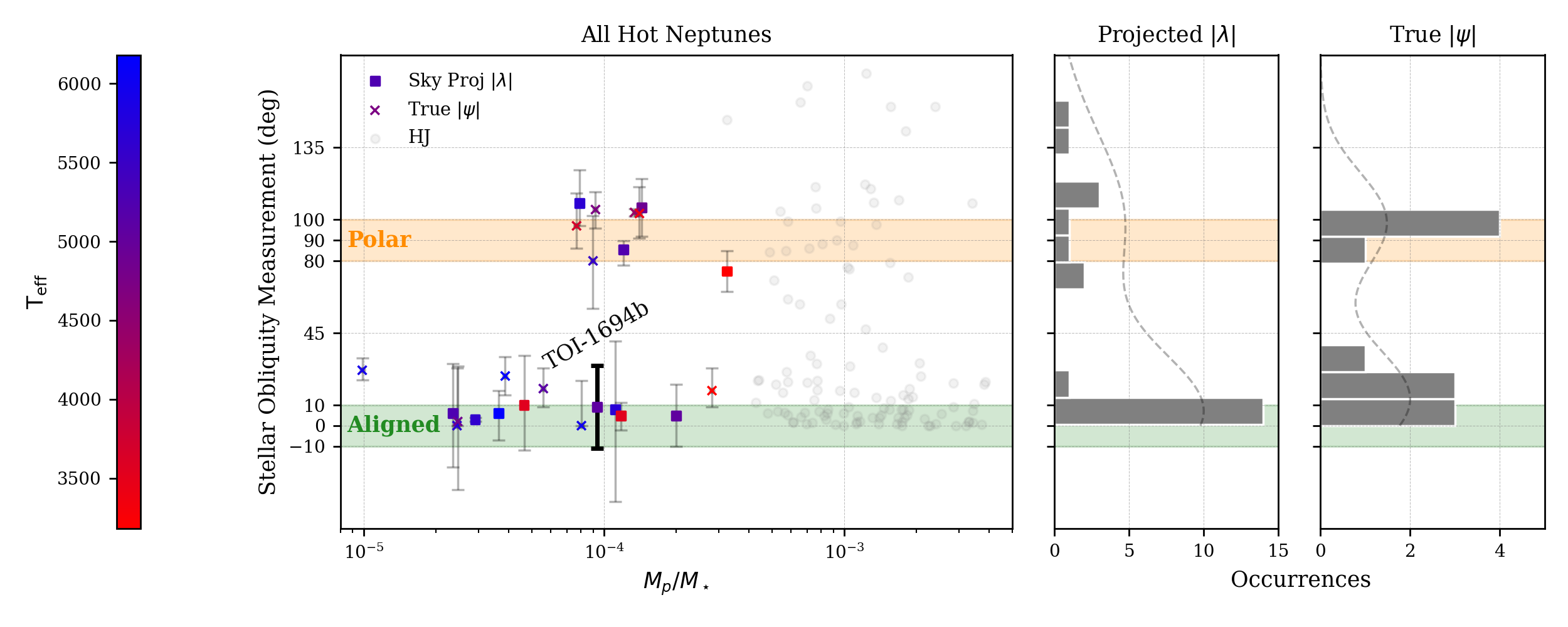}
    \caption{Absolute values of known obliquity measurements of hot Neptune systems from the catalog in \cite{southworth2011}, shown as a function of mass ratio and colored by stellar effective temperature (only cool stars have so far been explored). Cross shaped markers denote a true obliquity constraint which we preferred for this plot, and squares are used when only the sky-projected quantity was available. The regions of well aligned and polar orbits are shaded green and orange respectively to more clearly see the obliquity dichotomy. We include hot Jupiter systems as grey points for comparison. The histograms contain the entire suite of Neptune measurements we considered.}
    \label{fig:obliquities}
\end{figure*}

To place our measurement of TOI-1694b into context, we queried the TEPCat database \citep{southworth2011} for all the known spin-orbit angle measurements in conjunction with the NASA Exoplanet Archive\footnote{\url{https://exoplanetarchive.ipac.caltech.edu/}} for their planetary and stellar parameters. At the time of our query, there were roughly 220 unique systems, consisting mostly of hot Jupiters. We defined a hot Neptune as a planet with $2 R_\Earth \leq R_p$ and either $R_p \leq 6 R_\Earth$ or $M_p \leq 50 M_\Earth$, and with orbital period $P \leq 15$d. We plot the magnitude of those obliquity measurements in Figure \ref{fig:obliquities} as a function of mass ratio $M_p/M_\star$. For a few systems with only an upper limit for the mass, we used \cite{chenkipping2017} as an estimate. We required that the $1\sigma$ uncertainty on the chosen obliquity measurement be $\leq 40\degree$ to more clearly differentiate between aligned and polar systems. Including TOI-1694b, our dataset consisted of 12 values of $\psi$ and 23 values of $\lambda$ across 24 systems.

Figure \ref{fig:obliquities} indicates that the obliquities of Hot Neptune planets are bifurcated between nearly aligned orbits ($\psi,\lambda \lesssim 10\degree$) and nearly polar orbits ($\psi,\lambda \sim 90\degree$). To get an idea of the statistical significance of the observed dichotomy, we performed a Hartigan Dip Test (HDS, \cite{hartigan1985}) on the distributions of both \cospsi\ and \coslam. HDS assesses the distance between an empirical distribution and the nearest unimodal distribution, quantifying the evidence that an empirical dataset can be better modeled with a multi-modal (in this case, aligned and polar) distribution. We preferred the HDS as opposed to evaluating the Bimodality Coefficient because it is less prone to Type 1 errors \citep{kang2019}. The datasets are small, so we compute the p-values of the HDS using interpolation of a critical value table as implemented in the \textsc{diptest}\footnote{\url{https://github.com/alimuldal/diptest}} python package.

For the case of the true obliquities \cospsi, we find evidence of an aligned and polar population, with the p-value of a unimodal distribution being 0.018. We still caution the reader due to the small sample size -- p-values may not be robust in this regime. For sky-projected angles \coslam, the HDS evidence disappears with unimodal p-value of 0.43, which is expected given the smearing of projection effects. This suggests that an independent analysis of sky-projected angles may not be representative of the underlying bifurcated population modeled by \cospsi\ without constraints on the stellar inclinations. \cite{espinozaretamal2024} instead used a hierarchical Bayesian model (see, e.g., \cite{hogg2010}) in the framework of \cite{dong2023} to circumvent the need for stellar inclinations for every system. By applying this model to Hot Neptunes, they showed there is evidence for a polar population at the 1.5$\sigma$ level, but only when they assumed an informative prior. \cite{dong2023} noted that such models become robust to prior selection only when the sample size is $\gtrsim$ 50 planets, indicating that hot Neptunes are not yet well suited for hierarchical models.

For comparison, we also performed HDS on hot Jupiter obliquities around cool stars ($N\sim100$). We found the HDS p-value for the distribution of \cospsi to be 0.42, indicating little evidence for a distinct population of polar hot Jupiters. This is in agreement with the work of \cite{siegel2023}, \cite{dong2023}, and \cite{espinozaretamal2024}, who all pointed out that an aligned/polar multi-modality does not extend to hot Jupiter systems.

\subsection{Risk Factors for Polar Orbits}

By plotting Figure \ref{fig:obliquities} as a function of the mass ratio, we see that the polar hot Neptunes are clustered around a mass ratio of $M_p/M_\star \sim 10^{-4}$, and none are found below the cluster. The mass ratio is an inherently dynamical quantity which impacts the timescales of tidal circularization, tidal realignment, and migration. In the case of hot Jupiters around cool stars, higher mass ratio planets ($M_p/M_\star \sim 10^{-3}$) tend to be in more aligned orbits (\cite{hebrard2011}, \cite{albrecht2022}) than those at lower ratios. This follows the general logic that alignment probability for cool stars comes as a consequence of the realignment timescale, Equation \ref{eq:synctime}. For the small planets, however, the circumstances appear to be the exact opposite. Lower mass ratio systems, with longer alignment timescales, demonstrate consistent alignment across the board. It may be possible that the processes which generate polar orbits have a preferential mass ratio regime. It may also suggest that the realignment mechanism in \cite{zahn2008} is inadequate in explaining the distribution of observed obliquities.

We also note that polar orbits almost exclusively exist for planets with $P \le 6$~days. The only exception, HD~89345b \citep{bourrier2023}, is a peculiar case because it orbits a star currently evolving off the main sequence \citep{yu2018}. However, this does not appear to be due to the dependence of Equation \ref{eq:synctime} on the tidal parameter $a/R_\star$. While the polar systems have an average $a/R_\star = 15.7$, the aligned systems have a higher value of $16.8$. 


\subsection{Dynamical Sculpting of the Dichotomy} \label{sec:demographicdynamics}

The shape of the obliquity distribution for Hot Neptunes in Figure \ref{fig:obliquities} can give some insight as to which of the models in Section \ref{sec:dynamics} is the most relevant. The clustering at aligned and polar configurations is inconsistent with the predictions of a ZKL dominated theory. \cite{fabrycky2007} showed that ZKL oscillations result in a pileup of inclinations near the critical angles of $40\degree$ and $140\degree$, which are essentially uninhabited in Figure \ref{fig:obliquities}. Furthermore, Equation \ref{eq:kozaigr} suggests that polar orbits triggered by ZKL should be at longer periods where GR does not dominate apsidal precession, which is inconsistent with the absence of polar orbits above 6 days. 

If instead nodal precession were the dominant mechanism for generating Hot Neptune obliquities, we would observe a distribution like that illustrated in \cite{rubenzahl2021}, Figure 6. Depending on the assumed distribution of mutual inclinations, the probability density may have maxima at both aligned and polar configurations, however, we should still see a population of orbits at moderate inclinations ($\psi \in [30\degree-60\degree]$) if the typical amplitude is a polar configuration. Given the absence of Hot Neptune systems at moderate obliquities in Figure \ref{fig:obliquities}, the data does not strongly favor the sweeping of angles suggested by nodal precession. Furthermore, it would require that highly oblique Hot Neptunes all coexist with giant companions at moderately high obliquities themselves.

As demonstrated in Section \ref{sec:dynamics} for TOI-1694b, disk resonance sweeping preferentially creates polar orbits for GR dominated systems, even with only mild initial mutual inclinations of a few degrees for an outer giant planet. In the cases of a strong stellar quadrupole during disk dispersal or a long adiabatic time due to the outer planet being closely aligned, mild obliquities between $0-40\degree$ are predicted, which is currently in agreement with the observations. However, this model does not explain the pileup at high mass ratios because the resonance crossing is independent of the inner planets mass. Furthermore, it requires a contrived initial architecture with a Jovian planet and transition disk accompanying every polar Neptune. Only two polar systems, WASP-107 \citep{Piaulet2021} and HAT-P-11 \citep{yee2018}, have confirmed outer giants. Other polar hosts such as WASP-139 and HATS-38 even feature dedicated $\sim$decade long RV baselines which rule out the presence of Jovians to 10 au or more \citep{espinozaretamal2024}.

\section{Conclusion} \label{sec:conclusion}

In this work, we refined the transit ephemeris of TOI-1694b with TESS and modeled the RM effect with KPF, finding the sky-projected obliquity to be $\lambda={9\degree}^{+22\degree}_{-18\degree}$. We considered TOI-1694b, as well as the greater population of hot Neptunes with measured obliquities, within the context of several dynamical processes. ZKL oscillations are likely not the dominant excitation mechanism given that polar systems orbit in short periods where ZKL is ineffective due to GR effects. We demonstrated evidence of an independent polar population of Neptunes with a p-value of 0.018, and argue that such a distribution is not well explained by nodal precession. Early resonance encounters induced by the dispersing protoplanetary disk provide a plausible mechanism to generate polar orbits, but require the presence of giant companions, which are absent for at least some of these systems.

To explain the obliquities of Hot Neptunes, there are several challenges that a complete theory must address. It must predict the preferential sorting of Neptunes into polar and aligned configurations at short periods of $\lesssim$~6 days, across a diversity of outer system architectures. The suggested mechanism should operate effectively at mass ratios of 10$^{-4}$, but ineffectively at lower masses. Measuring more obliquities of small planets will be critical to further constrain such a peculiar theory, which may significantly impact on our understanding of planet formation as a whole.

\section{Acknowledgments}

We are grateful to Luke Bouma for valuable conversations about TESS photometry, and to Konstantin Batygin for a review of our dynamical arguments. A.W.H.\ acknowledges funding support from NASA award 80NSSC24K0161. This research was carried out, in part, at the Jet Propulsion Laboratory and the California Institute of Technology under a contract with the National Aeronautics and Space Administration and funded through the President's and Director's Research \& Development Fund Program.

The authors acknowledge the use of public TESS data from pipelines at the TESS Science Office and at the TESS Science Processing Operations Center at NASA Ames Research Center. This research has made use of the NASA Exoplanet Archive and Exoplanet Follow-up Observation Program website, which are operated by the California Institute of Technology, under contract with the National Aeronautics and Space Administration under the Exoplanet Exploration Program.

Some of the data presented herein were obtained at the W. M. Keck Observatory, which is operated as a scientific partnership among the California Institute of Technology, the University of California and the National Aeronautics and Space Administration. We wish to recognize and acknowledge the very significant cultural role and reverence that the summit of Maunakea has always had within the indigenous Hawaiian community. We are most fortunate to have the opportunity to conduct observations from this mountain.

\vspace{5mm}
\facilities{TESS, Keck I (KPF)}

\software{\textsc{astropy} \citep{astropy}, 
          \textsc{batman} \citep{batman}, 
          \textsc{emcee} \citep{emcee}, 
          \textsc{lightkurve} \citep{lightkurve}, 
          \textsc{matplotlib} \citep{matplotlib}, 
          \textsc{numpy} \citep{numpy}, 
          \textsc{pandas} \citep{pandas}, 
          \textsc{scipy} \citep{scipy}, 
          \textsc{spinspotter} \citep{holcomb2022}, 
          \textsc{tesscut} \citep{tesscut}, 
          \textsc{unpopular} \citep{unpopular}}

\appendix

\section{Ground Based Followup with Dragonfly} \label{sec:dragonfly}

For the sake of completeness, we wish to note that we also analyzed a ground-based followup observation using the Dragonfly lenslet array \citep{abraham2014}, conducted in January 2021 during another transit of TOI-1694b, obtained by user Christopher Mann and accessed from exofop.ipac.caltech.edu. However, for reasons we develop here, we chose to neglect this data and modeled only the TESS photometry.

The time series of these observations lasted roughly 5.3 hours, with 2.5 hours of out of transit baseline. Exposures were taken in the SDSSr band at 70s cadence across 24 different lenses. A time variable aperture function was used to model the evolving Point Spread Function (PSF). The typical aperture radius was 7 pixels (20") which is sufficiently small to neglect contamination from other Gaia sources. We initially attempted to model this data in conjunction with other 3 TESS Sectors. There were multiple factors which questioned the reliability of the photometry.

First, the flux errors reported by the data reduction routine were clearly underestimated by a large margin. Later observations taken at larger airmass had this problem exacerbated. We used an airmass-dependent error inflation scheme to better estimate the uncertainties across different phases of the transit. For the condition $\chi^2_\nu \approx 1$ to hold at all times during a fit of this data, the inflation ranged from roughly 2 times to as much as 7 times the original error across the time series. 

Second, the depth of the transit appears deeper in the Dragonfly photometry when compared to the TESS model. To assess the statistical significance of this difference, we performed a second fit of the Dragonfly data but allowed the depth parameter $R_p/R_\star$ to vary freely. After using an MCMC to estimate the uncertainties, we found that the detected $R_p/R_\star$ was 0.070, while the true TESS-only value of 0.061 was over 4$\sigma$ away based on the Dragonfly posterior.

\bibliography{main}{}
\bibliographystyle{aasjournal}



\end{document}